\documentclass[fleqn,usenatbib]{mnras}

\usepackage[T1]{fontenc}

\usepackage{graphicx}	
\usepackage{amsmath}	
\usepackage{url,times,graphicx,hyperref,amsmath,amsfonts,amssymb,aas_macros,color,epsfig,epstopdf}
\usepackage{booktabs}
\usepackage[utf8]{inputenc}
\usepackage{ae,aecompl}
\usepackage{subcaption}

\title[Tucana dSph: a distant backsplash galaxy of M31?]{The Tucana dwarf spheroidal: a distant backsplash galaxy of M31?}

\author[Santos-Santos, Navarro \& McConnachie]{
Isabel M.E.  Santos-Santos,$^{1,2}$\thanks{E-mail: isabel.santos@durham.ac.uk}
Julio F.  Navarro,$^{2}$
Alan McConnachie$^{3,2}$
\\
$^{1}$Institute for Computational Cosmology, Department of Physics, Durham University, South Road, Durham, DH1 3LE, UK\\
$^{2}$Department of Physics and Astronomy, University of Victoria, Victoria, BC V8P 5C2, Canada\\
$^{3}$NRC Herzberg Astronomy and Astrophysics, 5071 West Saanich Road, Victoria, BC, V9E 2E7, Canada
}

\date{Accepted XXX. Received YYY; in original form ZZZ}

\pubyear{2022}

\begin{document}
\label{firstpage}
\pagerange{\pageref{firstpage}--\pageref{lastpage}}
\maketitle

\begin{abstract}
    We use the APOSTLE 
  Local Group (LG) cosmological hydro-simulations to examine
  the properties of ``backsplash'' galaxies, i.e, dwarfs which were
   within the virial boundaries of the Milky Way (MW) or M31 in
  the past, but are today outside their virial radius
  ($r_{200}$). More than half of all dwarfs between
$1-2\,r_{200}$
of each primary are backsplash. More distant
  backsplash systems, i.e., those reaching distances well beyond
  $2\,r_{200}$, are typically close to apocentre of nearly
  radial orbits, and, therefore, essentially at rest relative to their
  primary. We use this result to investigate which LG dwarfs beyond
  $\sim500$ kpc of either primary could be a distant backsplash
  satellite of MW or M31. 
  Tucana dSph,
one of the few known quiescent LG field dwarfs, at
  $d_{\rm M31}\approx1350$ kpc and $d_{\rm MW}\approx880$ kpc, is a
  promising candidate. Tucana's radial velocity is consistent with
  being at rest relative to M31. Further, Tucana is located close to
  M33's orbital plane around M31, and simple orbit integrations
  indicate that Tucana may have been ejected during an early
  pericentric passage of M33 $\sim11$ Gyr ago, a timing which
  approximately coincides with Tucana's last episode of star
  formation. We suggest that Tucana may have been an early-infalling satellite of M31
  or M33, providing a compelling explanation for its puzzling lack of
  gas and ongoing star formation despite its isolated nature.
In this scenario, M33 should have
    completed some orbits around M31, a result that may help to
    explain the relative dearth of M33 satellite-candidates identified
    so far.
\end{abstract}




\section{Introduction}\label{sec:intro}

In the Lambda Cold Dark Matter (LCDM) cosmological paradigm galaxies
form at the centre of dark matter halos which grow 
hierarchically,  continuously accreting  smaller systems. Of all
accreted systems, the most
massive ones quickly spiral to the centre and merge with
the main halo, but lower mass systems may remain in orbit for a long time
and are today identified with satellite galaxies \citep[see;
e.g.,][and references therein]{Wang2011}.

Satellites are strongly affected by their host, both gravitationally,
as tides gradually pull away matter, and hydrodynamically, as the
circumgalactic gas of the primary ram-pressure strips away the gaseous
envelopes of subhalos, depriving them of star formation fuel and
eventually extinguishing their star formation activity \citep{Tolstoy2009}. 

This scenario leads naturally to differences between the properties of
satellite galaxies compared with dwarf
galaxies of similar mass in the field. In particular, it successfully
explains the origin of the environmental dependence of dwarf galaxy
types in the Local Group: the majority of satellites are quiescent, gas-free dwarf spheroidal
systems whereas field dwarfs are typically gas-rich dwarf irregulars
with ongoing star formation \citep[see; e.g.,][]{Grebel1998,Weisz2014}.

Given the importance of these environment-driven processes, it is
important to establish how far away from a galaxy they may
operate. Early work on galaxy clusters led to the realization that
environmental effects may extend well beyond the nominal virial boundary of
a system \citep{Balogh2000}, conventionally defined as the radius\footnote{More
  precisely, the virial radius is defined as the radius where the mean
enclosed density equals $200\times$ the critical density for
closure. We shall use the subscript ``200'' to identify quantities
measured at or within the virial boundary.}, $r_{200}$, where the circular
orbit timescale is comparable to the age of the Universe.

The reasons for the unexpectedly large ``radius of influence'' of a
primary system on its associated subsystems is twofold. One reason is that
many subhalo orbits are fairly radial, and  may reach
outside the virial radius during their first trip to apocentre after
accretion \citep{Mamon2004,Gill2005,Knebe2011a}. Indeed, most subhalos
first accreted $2$-$3$ Gyr ago into a halo like that of the Milky Way
are expected to be at present outside the virial radius \citep{Barber2014}. These so-called
``backsplash'' galaxies are especially abundant just outside the virial
radius, representing a fraction of that may exceed $\sim 50$\% of subhalos with
$1<d/r_{200}<2.5$
\citep{Garrison-Kimmel2014,Buck2019,Simpson2018,Applebaum2021,Bakels2021}.

The second reason is that many subhalos come as members of virialized
groups which are tidally dissociated soon after first infall into the
primary halo. As discussed in detail by \cite{Sales2007} and
\cite{Ludlow2009}, some subhalos may gain enough energy during the
disruption of their group to be expelled much further away, to
distances as far as 5 virial radii or beyond \citep[see also][]{Teyssier2012}. Systems on these extreme
orbits are typically a small fraction of the low-mass members of the group,
whose heavier members typically stay tightly bound to the primary. Identifying
these ``extreme backsplash'' cases therefore requires not only some
evidence for dynamical association, but also the existence of a more
massive ``parent'' progenitor to help propel them into highly energetic orbits.

In the cosmological context of the Local Group, the above discussion
suggests the existence of a rare population of low-mass field dwarfs,
located far away (out to $\sim 1.5$ Mpc) from the Milky Way (MW) and
Andromeda (M31), but showing properties consistent with 
satellites of either of them, such as lack of ongoing star formation.
These galaxies are indeed unusual,  since most isolated
galaxies discovered so far in the Local Group field and beyond are
currently star-forming \citep{Geha2012}.

To date, the only known examples of field dSph galaxies within $1.5$
Mpc of the LG midpoint are Cetus, Tucana, and And XVIII, currently at
$\sim 755 (674)$, $\sim 877 (1345)$, and $\sim 1330 (580)$ kpc from the
MW (M31), respectively \citep{Whiting1999,Lavery1992,McConnachie2008}.
All three show little to no gas content and predominantly old
stellar populations formed roughly $\sim 9$-$10$ Gyrs ago
\citep{Castellani1996,Monelli2010a,Monelli2010b,Savino2019,Makarova2017}.
Further away, at $\sim 2$ Mpc, the only other examples of quiescent
dwarfs known are KKR25 and KKs3
\citep{Karachentsev2015,Karachentsev2001}, plus the recent discoveries
of Tucana B \citep{Sand2022} and COSMOS-dw1 in the COSMOS-CANDELS
field beyond the LG \citep{Polzin2021}.

The origin of isolated dwarf galaxies with no
recent star formation activity remains poorly understood, but it has been argued
that, in the case of Cetus and Tucana, they may have resulted from
either a backsplash interaction with the MW \citep[e.g.][]{Sales2007,Fraternali2009,Teyssier2012} or from
ram-pressure stripping with the cosmic web
\citep[e.g.][]{BenitezLlambay2013}. More recently, a novel proposal
associating them with the effects of the photoionizing background has
been put forward by \citet{PereiraWilson2022}.

We use in this paper distant backsplash dwarfs in the APOSTLE
cosmological hydrodynamical simulations \citep{Sawala2016,
  Fattahi2016} to characterize the kinematic properties of such
systems in the Local Group. We focus on seemingly isolated galaxies at
distances larger than $\sim 500$ kpc from the MW and M31, noting as
well that, as reported in earlier work, many dwarfs between
$r_{200}<d<2.5\, r_{200}$ (roughly out to $\sim 500$ kpc of the MW or
M31) are indeed backsplash galaxies.


This paper is organised as follows.  In Sec.~\ref{sec:methods} we
describe the APOSTLE simulations and the observational data used.  Our
results on distant backplash galaxies in APOSTLE are presented in
Sec.~\ref{sec:apostle}.  Section~\ref{sec:bsLG} shows our analysis of
Local Group dwarfs in light of the simulation results.
Finally in Sec.~\ref{sec:tuc} we focus on the Tucana dSph and provide
evidence supporting a hypothetical backsplash
origin. Sec.~\ref{sec:conclu} summarizes our conclusions.

\section{Methods}\label{sec:methods}

\subsection{Numerical simulations}\label{sec:sims}

The APOSTLE simulations are a set of cosmological volumes chosen to
include two massive primary halos with masses, relative distance, relative
radial velocity and surrounding Hubble flow similar to that observed
for the Milky Way (MW) and M31 pair \citep{Fattahi2016}.  In this work
we have used 4 volumes run at the highest resolution in APOSTLE
(labelled ‘L1' level in previous literature).  These runs have initial dark
matter and gas particle masses of $m_{\rm DM}\sim 5\times 10^4$
M$_\odot$ and $m_{\rm gas}\sim 1\times 10^4$ M$_\odot$, respectively,
and a gravitational softening length of $134$ pc at $z = 0$.  The
zoom-in region of each APOSTLE volume fully contains a sphere of
radius $r \sim 3.5$ Mpc from the midpoint of the MW and M31 ``primary''
halos.

APOSTLE used the EAGLE galaxy formation code
\citep{Schaye2015,Crain2015}. This model includes subgrid physics
prescriptions for star formation in gas exceeding a
metallicity-dependent density threshold, radiative cooling of gas,
stellar feedback (from stellar winds, radiation pressure and
supernovae), homogeneous X-ray/UV background radiation, supermassive
black hole growth and AGN feedback (the latter have negligible effects
on dwarf galaxies).

APOSTLE assumes a flat $\Lambda$CDM cosmological model following
WMAP-7 parameters \citep{Komatsu2011}: $\Omega_{\rm m}=0.272$;  $\Omega_{\Lambda} = 0.728$; $\Omega_{\rm bar} = 0.0455$;  $H_0 = 100\, h$ km s$^{-1}$ Mpc$^{-1}$; $\sigma  = 0.81$; $h = 0.704$.

\subsubsection{Simulated galaxies}

Haloes and subhaloes in APOSTLE have been identified using the
friends-of-friends (FoF) groupfinding algorithm \citep{Davis1985}
(with linking length equal to $0.2$ times the mean interparticle
separation) and the SUBFIND halo finder
\citep{Springel2001,Dolag2009}.

Simulated galaxies are halos where star formation has led to the
formation of a luminous component. In APOSTLE, this restricts galaxy
formation to field halos more massive than $M_{200}\sim 10^9\,
M_\odot$. Satellite galaxies may exist in subhalos with lower mass,
because of tidal stripping; see for details
\citet{Fattahi2018}.

We shall define galaxies \textit{associated} with each APOSTLE primary
as those that have been within the virial radius of the primary's most
massive progenitor at some time during its evolution.  Associated
galaxies include satellites (i.e, galaxies within $r_{200}$ at $z=0$)
and backsplash galaxies (i.e, associated galaxies located today
outside the virial radius of the primary). Backsplash systems were
identified by tracking back in time all galaxies found outside the
virial radius of both main primaries at $z=0$\footnote{In this work we
  do not distinguish galaxies which are today satellites of one of the
  primaries, but were associated with the other primary at an earlier time \citep[see,
  e.g.][]{Newton2021}.}.
Each of the main APOSTLE primaries present halo masses $M_{200}$ ranging from 
  $0.78$ to $2.05\times 10^{12}$ M$_\odot$, with primary-secondary mass ratios in the range $\sim 0.33$-$0.96$.


\subsection{Observational data}\label{sec:data}

In this work we consider the currently known Local Group dwarf galaxies
within $\sim 1.5$ Mpc of the midpoint between the MW and M31.
We use the latest 
position and velocity
data in the \citet{McConnachie2012} compilation of nearby galaxies\footnote{see \url{https://www.cadc-ccda.hia-iha.nrc-cnrc.gc.ca/en/community/nearby/}, and references therein.}.
We refer to dwarf galaxies within 300 kpc of the MW or M31 as
``satellites'' of that primary; the rest are considered ``isolated'' or ``field'' dwarfs.

For M31 and its satellite M33 (i.e., Triangulum) we adopt the
positions and velocities derived from the combined Gaia DR2 and
\textit{HST} proper motions by \citet{vanderMarel2019}.

Galactocentric positions and velocities have been computed assuming a
Galactocentric distance for the Sun of $R_\odot=8.29$ kpc, a peculiar
velocity with respect to the LSR of
$(U_\odot,V_\odot,W_\odot) = (11.1, 12.24,7.25)$ km/s
\citep{Schonrich2010}, and a circular velocity for the local standard
of rest (LSR) of $V_0=239$ km/s \citep{McMillan2011}.

We make use of  updated Gaia EDR3 systemic proper motions 
for a set of distant Local Group dwarf galaxies for which such data
has been measured \citep{McConnachie2021}.
For dwarfs without proper motion measurements we 
convert the heliocentric line-of-sight velocities to the Galactic standard of rest (GSR) as
$
\vec{V}_{\rm GSR}= \vec{V}_{\rm hel} + \vec{V}_{\odot,\rm proj}
$;
where $\vec{V}_{\odot,\rm proj}$
is the projection of the Sun's motion ($\vec{V}_{0}+\vec{V}_{\rm pec}$) along the Galactocentric radial direction to the dwarf galaxy.

Tables~\ref{tab:m31data} and \ref{tab:m31dataderived} present the
specific data values used for M31, M33 and Tucana, the objects which
we will focus on later in the paper (see Sec.~\ref{sec:tuc}).\footnote{See also \citet{Taibi2020,Savino2022}. Note that none of our conclusions are changed by using these alternative data values.}

\begin{table*}
\centering
\caption{Observational data used in this work for M31, M33 and Tucana.  Columns show right ascension and declination, distance from the Sun,  heliocentric line-of-sight velocity and proper motions. References: \citet{McConnachie2012,vanderMarel2012,vanderMarel2019}.
In red are our predicted proper motions for Tucana if it is a backsplash galaxy of M31, computed by assuming it is at rest with respect to M31 (see Sec.~\ref{sec:tuc}). }
\begin{tabular}{ l l l l l l l }
\toprule
Galaxy  & RA (deg) & Dec (deg) & $D_{\odot}$ (kpc) & $V_{\rm hel}$ (km/s) & $\mu_{\rm RA*}$ (mas/yr) & $\mu_{\rm Dec}$ (mas/yr)  \\
\midrule
M31 & 10.684 & 41.269 & 770$\pm$40 & -301$\pm$1 & 0.049$\pm$0.011 & -0.038$\pm$0.011 \\
M33 & 23.462 & 30.660 & 794$\pm$23 & -180$\pm$1 & 0.024$\pm$0.007 & 0.003$\pm$0.008\\
Tucana & 340.456 & -64.419 & 887$\pm$50 & 194$\pm$4.3 & \textcolor{red}{\textbf{0.0206}} & \textcolor{red}{\textbf{-0.0754}} \\
\bottomrule
\end{tabular}
\label{tab:m31data}
\end{table*}

\begin{table*}
\centering
\caption{Galactocentric position and velocities for M31, M33 and Tucana derived from data in Tab~\ref{tab:m31data}. }
\begin{tabular}{ l l l l l l l l}
\toprule
Galaxy & $X$ (kpc) & $Y$ (kpc) & $Z$ (kpc) & $V_{\rm rad}$ or $V_{\rm GSR}$ (km/s) & $V_X$ (km/s) & $V_Y$ (km/s) & $V_Z$ (km/s) \\
\midrule
M31 & -378.95 & 612.66 &  -283.12  & -108.91 & 34.99 & -123.82 & -17.02 \\
M33 & -476.09 & 491.06 & -412.86 & -35.17  & 44.34 & 90.95 & 125.10\\
Tucana & 470.99 & -652.71 & -362.36 &  91.42 & - & - & - \\
\bottomrule
\end{tabular}
\label{tab:m31dataderived}
\end{table*}


\section{Results}

\subsection{Backsplash galaxies in APOSTLE }\label{sec:apostle}

\begin{figure}
\centering
\includegraphics[width=\linewidth]{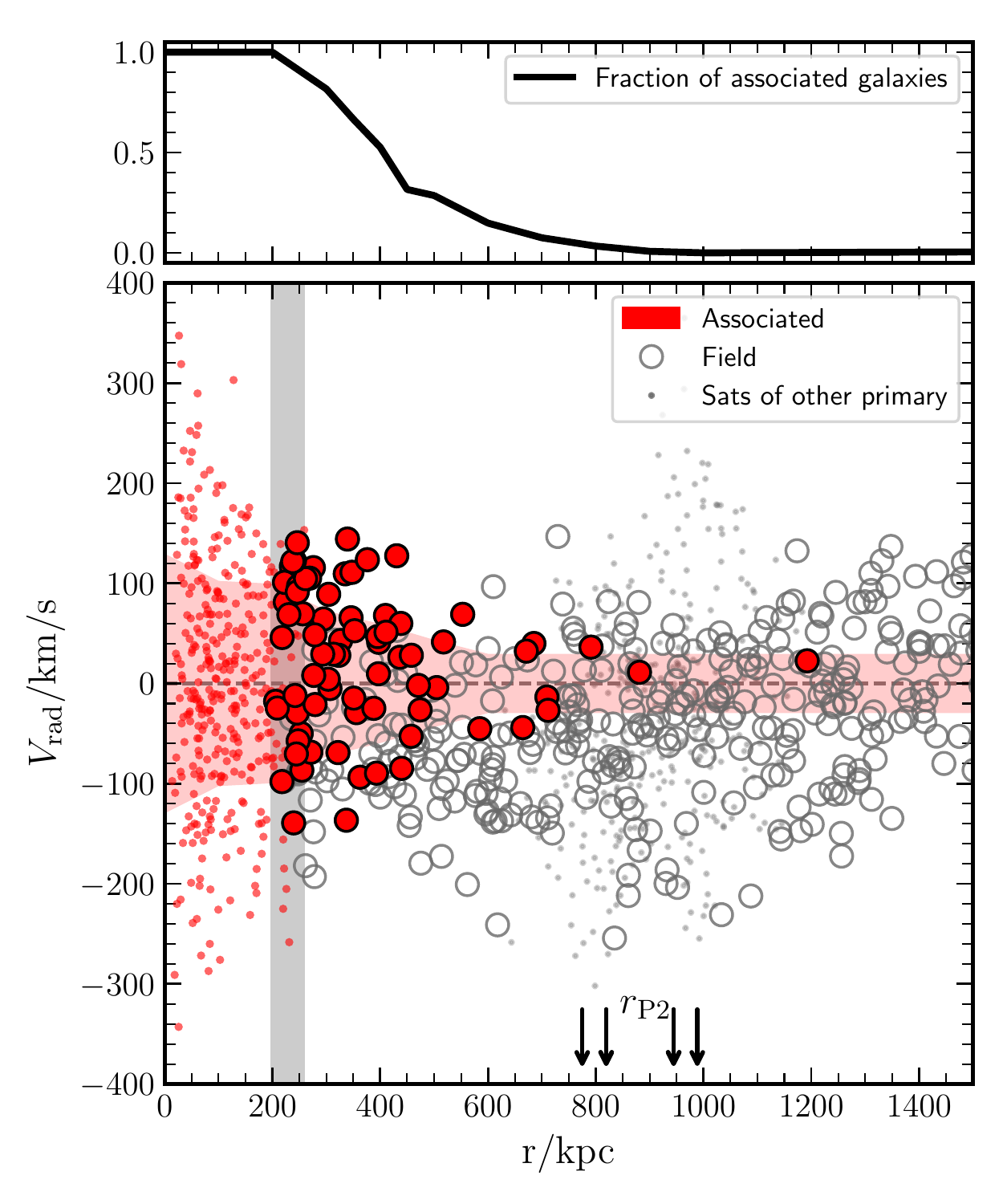}
\caption{Radial velocity vs distance for galaxies  in the APOSTLE Local Group
  simulations in the reference frame of one of its primaries. Results
  for all $8$ primaries are stacked.
 Associated galaxies are shown in red, being either satellites (i.e.,
 within $r_{200}$, smaller circles) or backsplash galaxies (i.e.,
 outside $r_{200}$, larger circles).
Field dwarfs not associated with the primary are shown as open gray circles.
Satellites of the other primary in the volume are shown as small gray dots.
For reference, a vertical gray band indicates the 10-90 percentile
range of $r_{200}$ values for all $8$ APOSTLE primaries.
The $\pm1\sigma$  radial velocity dispersion of associated galaxies as a function of distance is shown with a red shaded area. 
The radial distance to the second primary in each APOSTLE volume, $r_{\rm P2}$, is marked with an arrow for reference.
An upper auxiliary panel indicates the average fraction of associated
galaxies,  over all galaxies in the volume,  as a function of radial distance from the primary. 
}
\label{fig:vrad}
\end{figure}

Fig.~\ref{fig:vrad} shows the radial velocity versus distance from the
primary for all galaxies identified in the 4 APOSTLE high-resolution
volumes studied. This figure is centered on each of the 8 available
primaries and shows a stack of all luminous galaxies in each of the simulated
volumes.

``Associated'' galaxies are shown in red, being either satellites
(small circles) or backsplash galaxies (big circles with black
edges). A vertical shaded area delimits the 10-90 percentile range of
$r_{200}$ values for the 8 primaries, $196$-$261$ kpc , which
separates the overall satellite and backsplash populations.

``Isolated'' dwarfs are shown as gray open circles.  Note that because
of the binary nature of the Local Group, some of the isolated galaxies
could be backsplash galaxies of the other primary in the same
volume. For reference, the radial distance to the other primary,
$r_{\rm P2}$, is marked with an arrow.

We find an average of $\sim 43$ satellites and $\sim 9$ backsplash per
primary, down to a limit of one star particle, or roughly
$M_*\sim 10^4\, M_\odot$. This is best regarded as a lower limit, as
the raw number is very likely affected by numerical limitations.
 There are actually more associated subhalos outside than inside the
 virial radius \citep{Ludlow2009}, but the vast majority of them are
 low-mass subhalos without stars in them.

The upper panel in Fig.~\ref{fig:vrad} shows the fraction of associated galaxies over all dwarfs in the simulated Local Group, as a function of radial distance from the primary.  
In APOSTLE,  more than $>80\%$ ($50\%$) of dwarfs within $300$ $(400)$ kpc of a primary are associated to it, emphasizing that the virial radius
does not represent a true physical boundary separating objects that have or have not been influenced dynamically by the primary.
At $550$ kpc, only $25\%$ of dwarfs are associated; at $700$ kpc, fewer than $10\%$ are. 
The furthest backsplash case we find is at a distance of $\sim 1.2$
Mpc from its primary, roughly $6\times r_{200}$.

How can some backsplash galaxies reach such large distances ($\sim 1$
Mpc) from the primary?  As explained in \citet{Sales2007} and
developed further by \citet{Ludlow2009}, low-mass galaxies can be
ejected out to large distances during the tidal dissociation of groups
of dwarfs during their first infall. The tides induce the formation of
two ``tails'' of subhalos as the group disrupts; one that loses and
another one that gains orbital energy during disruption \citep[see;
e.g., Fig.~4 of][]{Ludlow2009}. The subhalos
carried away in the latter tail can sometimes reach very large
distances, and formally even ``escape'' the primary. Typically, the
lowest mass and least bound subhalos in the group are the ones more
susceptible to being propelled to extreme orbits.

As may be expected, distant associated galaxies are close to the
apocentre of their orbits, and are thus basically at rest with the
primary. This is seen in  Fig.~\ref{fig:vrad}, which shows that the
radial velocities of very distant backsplash systems, which are on
nearly radial orbits, decrease
systematically with increasing distance, approaching zero for the most
distant ones. (The red shaded region in Fig.~\ref{fig:vrad} shows the
radial velocity dispersion of associated systems as a function of
distance.)
On the other hand,  at similar distances, the radial velocities of
other, unassociated dwarfs in the Local Group span a wide range of
values.

This result suggests that a low relative velocity may in principle be
used as a robust criterion to identify candidate ``extreme
backsplash'' galaxies (i.e., those located at
$d>2$-$3\, r_{200}$) associated to a given
primary. We use this finding next to identify galaxies in
the Local Group that may have been previously associated to the
MW or M31.

\subsection{Distant backsplash candidates in the Local Group}
\label{sec:bsLG}

\begin{figure}
\centering
\includegraphics[width=\linewidth]{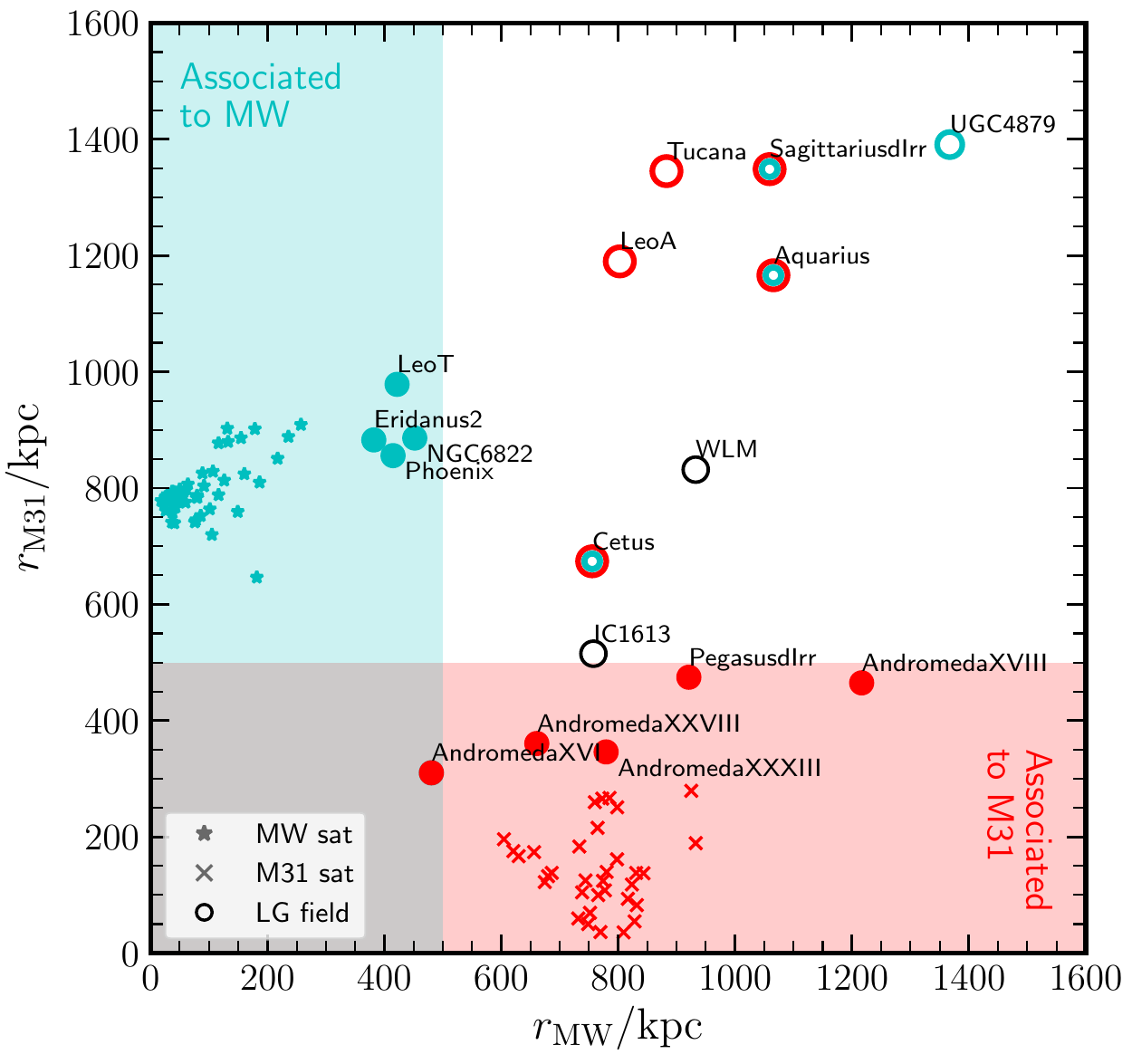}
\caption{Distances to MW and M31 for Local Group dwarfs within $1.5$
  Mpc of its midpoint.  MW satellites are shown with star symbols, M31
  satellites as 'x' symbols and field dwarfs as circles with labels.
  Objects within $500$ kpc of either primary are colored (cyan for the
  MW and red for M31).  Some field galaxies outside 500 kpc are
  colored as well, according to radial velocity criteria introduced in
  Fig.~\ref{fig:vradvpred} which identifies them as backsplash
  candidates of either primary.  }
\label{fig:rmwm31}
\end{figure}

\begin{figure*}
\centering
\includegraphics[width=0.6\linewidth]{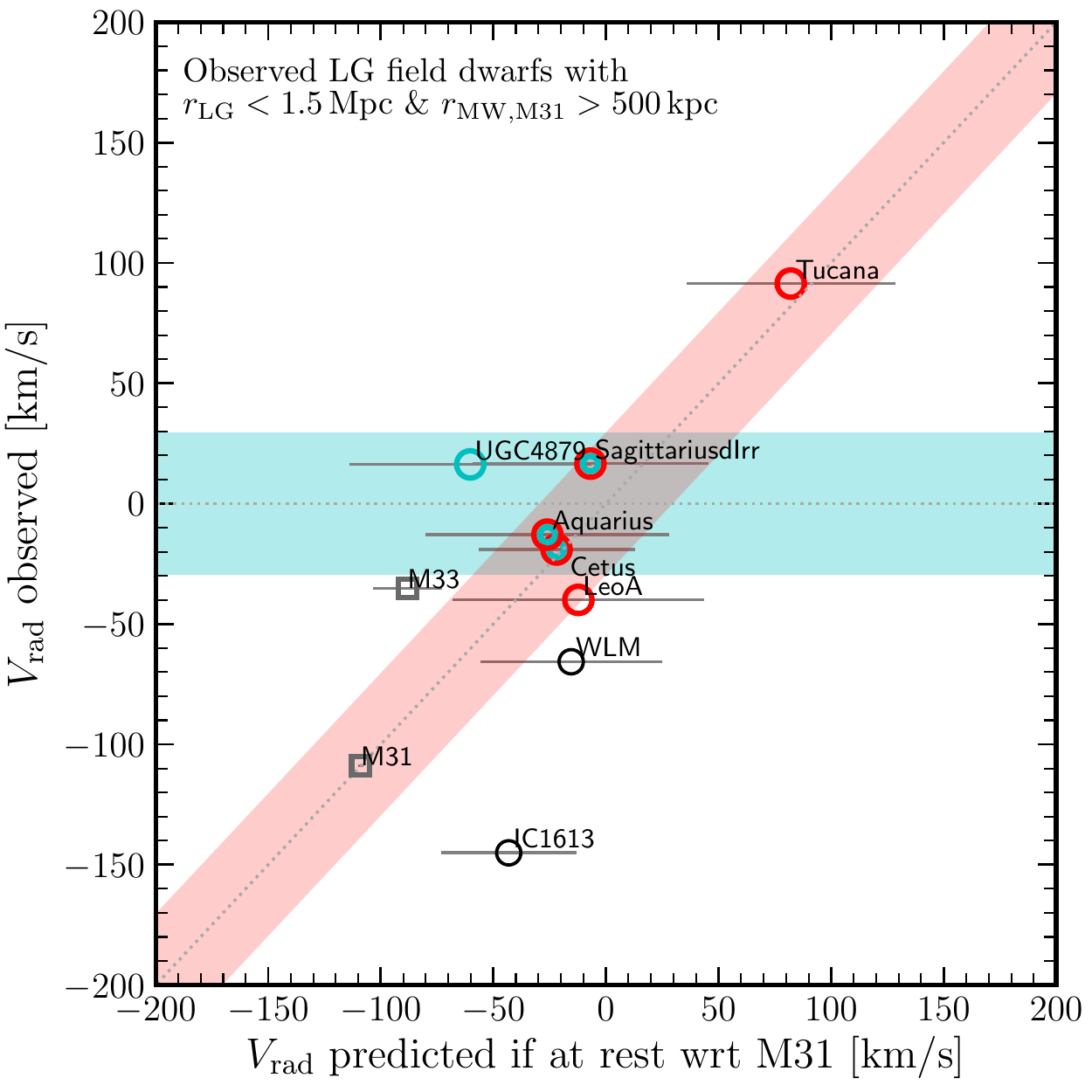}
\caption{Observed Galactocentric radial velocity vs that predicted if
  the galaxy was at rest with respect to M31.  Only Local Group field
  galaxies within $1.5$ Mpc from the Local Group's midpoint and
  outside $500$ kpc of the MW and M31 are considered. The cyan and red shaded bands mark an area
  of $\pm1\sigma$ in $V_{\rm rad}$, as measured for distant
  backsplash galaxies from the APOSTLE Local Group simulations (see
  Fig.\ref{fig:vrad}).  Galaxies within the horizontal cyan band
  present radial velocities compatible with being backsplash galaxies
  of the MW. Galaxies within the red diagonal band present radial
  velocities compatible with being backsplash galaxies of M31.  The
  dotted line marks the 1:1 correspondence.  Error bars indicate the
  minimum and maximum ``predicted'' $V_{\rm rad}$ values when
  considering the uncertainties in M31's proper motion data.  For
  reference, M33 and M31 are shown as gray squares.  }
\label{fig:vradvpred}
\end{figure*}

Figure~\ref{fig:rmwm31} shows the radial distance to the MW versus
the radial distance to M31,  for observed LG dwarf galaxies within
$\sim 1.5$ Mpc from the LG midpoint.

The shaded areas in Figure~\ref{fig:rmwm31} highlight distances within
$500$ kpc of the MW (cyan) or M31 (red). Objects
within these boundaries are likely associated to that primary, and are
colored accordingly.  In each case, this includes the
satellites (i.e., those with $d<300$ kpc, shown as star symbols for MW satellites or 'x' symbols for
M31 satellites) and dwarfs with $300<d$/kpc$<500$ which, according
to APOSTLE, have fairly high probability of being backsplash galaxies,
shown as circles. One galaxy, And XVI, overlaps both samples as it is
located at $r_{\rm MW}=450$ kpc and $r_{\rm M31}=310$ kpc. We assume
it is associated to M31, to which it is closer.


Eight dwarf galaxies are at larger distances (i.e.  Aquarius,
Cetus, IC1613, LeoA, Sagittarius dIrr, Tucana, UGC4879, WLM), and we
will consider them as potential distant backsplash candidates for the rest of
our study.

Any galaxy  from this subsample which is a backsplash of
the MW or of M31 should be essentially at rest relative to its primary.
We illustrate this idea in Fig.~\ref{fig:vradvpred}.
This figure shows the Galactocentric radial velocity
of each of these galaxies ($V_{\rm rad}$) versus the Galactocentric
radial velocity they would have \textit{if they were at rest relative
  to M31} ($V_{\rm pred}$). Note that we only use the radial velocity
component in this diagnostic because proper motions for most distant
dwarfs are unknown.

To compute $V_{\rm pred}$,
we simply assume that, relative to the MW, the 3D velocity vector of
the dwarf galaxy is the same as that of M31, and project accordingly.  $V_{\rm pred}$ for a
certain dwarf is thus calculated by projecting M31's Galactocentric 3D
velocity vector along the MW-dwarf radial direction as:
\begin{equation}
V_{\rm pred} = \frac  {   \vec{V}_{\rm M31,MW} \cdot  \vec{r}_{\rm dwf,MW}   }{|\vec{r}_{\rm dwf,MW}|}.
\end{equation}

A cyan shaded area indicates a region of $\pm1 \sigma_{\rm rad}$
around $V_{\rm rad}=0$ km/s in the y-axis, where
$\sigma_{\rm rad} =\pm 29$ km/s, the radial velocity dispersion of
distant ($d>500$ kpc) backsplash systems in APOSTLE (see
Fig.~\ref{fig:vrad}).

Dwarfs in the cyan area are compatible with
being backsplash galaxies of the MW and have been colored in cyan.
Alternatively, dwarfs fallling in the red shaded area around the 1:1
line --with a width also equal to $\pm1\sigma_{\rm rad}$-- have
observed radial velocities compatible with being backsplash galaxies
of M31 and are colored in red.

For reference, M31 and M33
(Triangulum) are shown as grey squares. M31 falls exactly on the 1:1 line
by construction.  Error bars correspond to the minimum and
maximum $V_{\rm pred}$ obtained when considering the uncertainties in
M31's proper motion data.
 
Six out of eight dwarfs are plausible backsplash candidates according
to this criterion.  UGC4879, Sagittarius dIrr, Aquarius and Cetus
could have been associated to the MW. The last three, plus possibly Leo
A, are also compatible with being backsplash candidates of M31. The
Tucana dSph, on the other hand, stands out as a clear M31 distant
backsplash candidate, with a Galactocentric radial velocity in very
close agreement with that expected for an object at rest relative to
M31.

\subsection{ Cetus and Tucana as distant backsplash candidates}
\label{sec:tuc}

The case of Cetus and Tucana as backsplash candidates are of
particular interest given that they are two of the few Local Group field
dSphs. Because of their low gas content, as well as
their predominantly old stellar populations, these systems resemble MW
or M31 satellites rather than field dwarfs
\citep{Monelli2010a,Monelli2010b,Fraternali2009}, and it is therefore
tempting to associate them with backsplash systems.

\begin{figure*}
\centering
\includegraphics[width=0.65\linewidth]{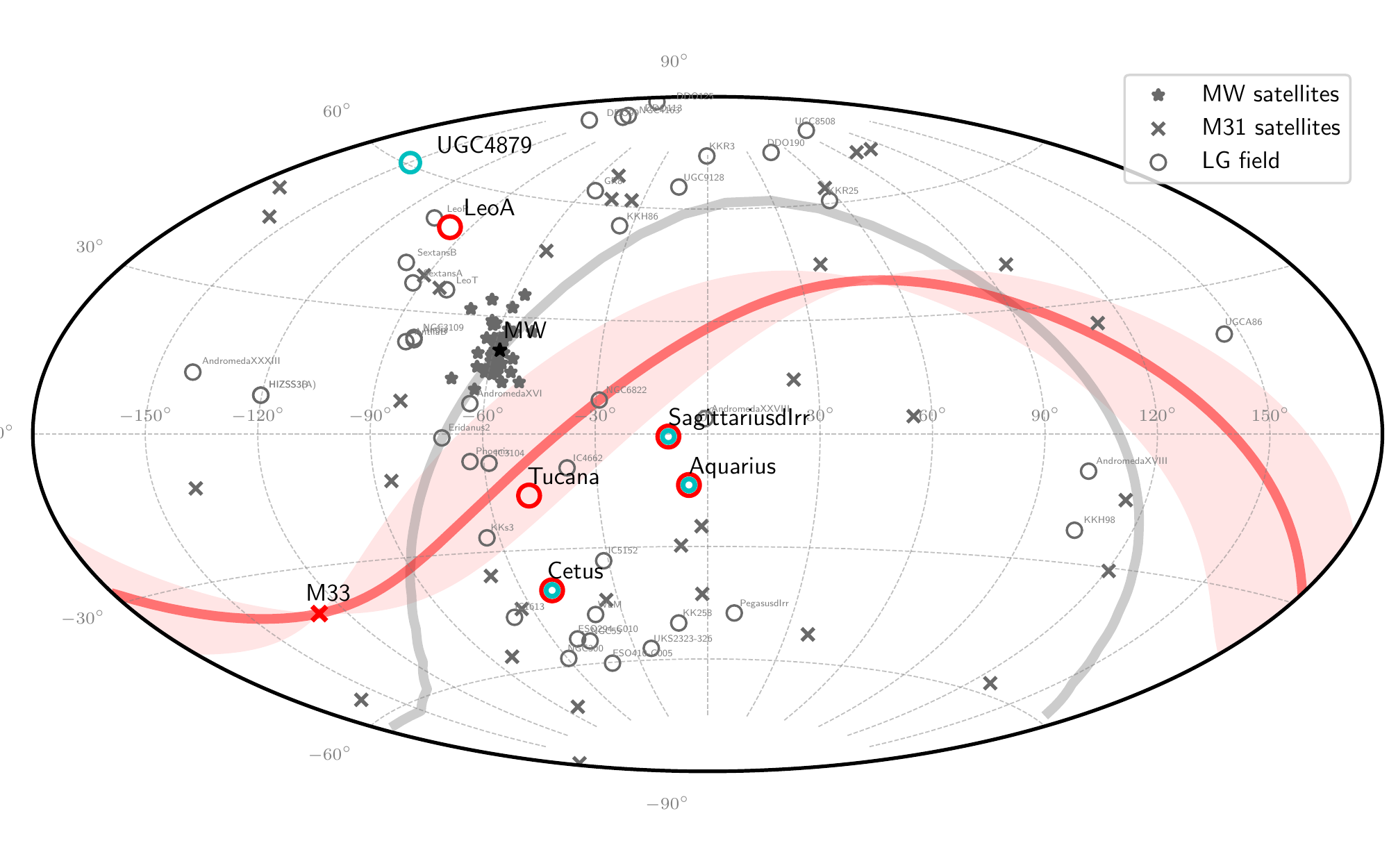}
\includegraphics[width=0.34\linewidth, trim=0 -2cm 0 0, clip]{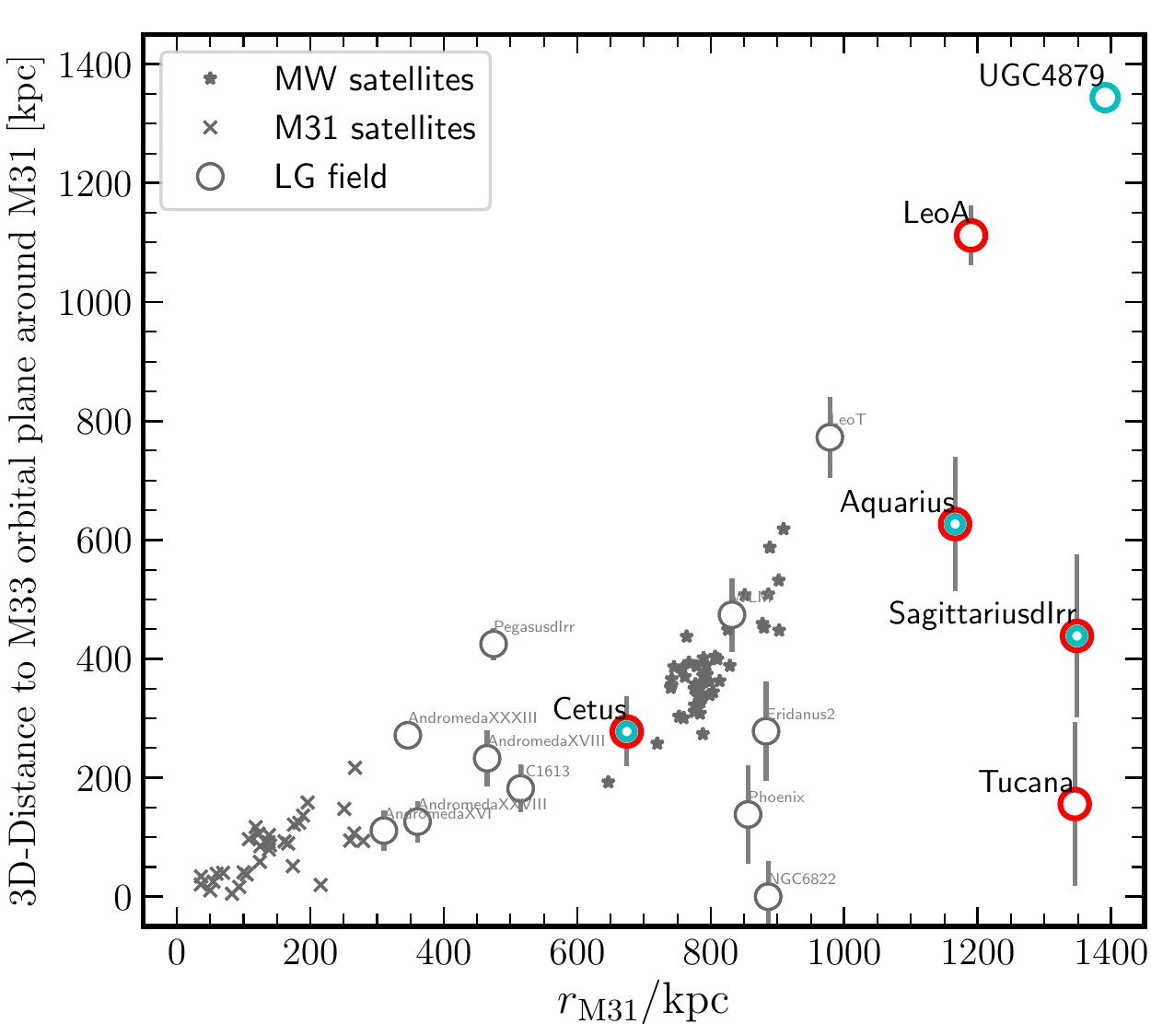}
\caption{ \textit{Left:} Aitoff sky projection of the Local Group
  dwarf galaxies in a reference system centred on M31 and oriented
  such that latitude $b=0^\circ$ is aligned with the MW's disk.  MW
  satellites are shown as star symbols, M31 satellites are shown as
  'x' symbols and field dwarfs within 1.4 Mpc of the Local Group are
  shown as circles.  Dwarfs shown in color are distant backsplash
  candidates of the MW (cyan) or M31 (red).  A thick grey line
  marks the MW's orbital plane, whereas a thick red line marks M33's orbital
  plane around M31.  A red shade indicates the uncertainty on
  this orbital plane as inferred from M33's proper motion errors.
  Tucana lies very close to M33's orbital plane.
  \textit{Right:} 3D-distances from Local Group dwarfs to the orbital
  plane of M33 around M31, vs their distances to M31.
  Errorbars show $\pm 1\sigma$ uncertainties in the distances to M33's
  orbital plane, computed by randomly sampling M33's proper motion,
  including errors.  }
\label{fig:fielddist}
\end{figure*}

Are there any other further hints that Tucana or Cetus may actually be
distant backsplash systems? Both seem to satisfy the low radial
velocity dispersion criterion (see Fig.~\ref{fig:vradvpred}), but so
do several other distant LG dwarfs. As discussed in
Sec.~\ref{sec:intro}, further evidence for a backsplash origin may
include the identification of a plausible ``parent'' satellite system
whose tidal dissolution may have expelled the dSph.  Both the MW and
M31 have satellites massive enough to be plausible parents of either
Tucana or Cetus; in particular, the Magellanic Clouds in the case of the MW and the
Triangulum galaxy (M33) in the case of M31.

For the Clouds, there is now robust evidence that they are just past
the first pericentric approach of their orbit around the MW
\citep{Besla2012,Kallivayalil2013}. This disfavours them as possible
parents of distant backsplash systems, as these objects are ejected
after a pericentric passage, and they would require several Gyr to
travel to their current location. A similar reasoning disfavours the
Sagittarius dSph as a potential parent, since the latest orbital
modeling suggests that Sagittarius first approach to the MW happened
only $\sim 5$-$6$ Gyr ago \citep{Laporte2018b}. As we shall see below,
reaching the large distances of Cetus and Tucana require that the
ejection must have occured much earlier than that.

It is in principle possible that a massive progenitor could have merged with the
central galaxy soon after pericentre, but there is little evidence
that the MW has undergone a substantial merger in the recent past. The
lack of an obvious parent system therefore suggests that none of the
MW distant backsplash candidates in Fig.~\ref{fig:vradvpred} (i.e.,
those in the cyan band) have actually been associated with the MW in
the past.

Could some of the distant candidates be associated with the accretion
of M33 into M31? Since proper motions and radial velocities are
available for both of these systems, it is possible to estimate the 3D relative
velocity of the M31-M33 pair using
the data compiled in Table~\ref{tab:m31data}. The resulting
velocity,  $V_{\rm M31 \textrm{-} M33}\sim 258$ km/s,  is not much
higher than the rotation speed of M31 
\citep[$V_{\rm max}\sim226$ km/s,][]{Carignan2006}
and likely well below the M31
escape velocity at M33's location. M33 is thus likely to be
on a fairly bound orbit
and may have completed a few pericentric passages in the past 
\citep[see; e.g.,][]{McConnachie2009,Patel2017,vanderMarel2019}, 
making it a plausible ``parent'' for backsplash systems. 

We investigate further a possible connection between the distant LG
dwarfs and the M31-M33 pair in Fig.~\ref{fig:fielddist}, where we
show, in an Aitoff projection, the position of various LG galaxies in
an {\it M31-centric} reference frame. We choose the ``equatorial
plane'' of the projection ($b=0^\circ$) to coincide with the MW
plane, and the N-S direction of the polar axis so that MW is in the
northern hemisphere of the projection. The MW-M31 orbital plane is
shown by the thick grey curve in Fig.~\ref{fig:fielddist}; the M33
orbital plane around M31, on the other hand, is shown by the thick red
curve.

The latter plane is especially significant, since we would expect that
systems that may have been expelled during the accretion of M33 into
M31 to share the same orbital plane of the main progenitor and to
remain close to it after ejection \citep[see;
e.g.,][]{Sales2011,Santos-Santos2021}. This reasoning singles out the
Tucana dSph in Fig.~\ref{fig:fielddist} as the most promising
candidate of them all. Indeed, Tucana is only $6.6^\circ$ ($<150$ kpc)
away from
the M33 orbital plane, which is only about a
tenth of its current distance from M31 (see right-hand panel of
Fig.~\ref{fig:fielddist}).

This could be, of course, just an extraordinary coincidence, but it
motivates us to examine further a potential association between
Tucana and M33/M31. A powerful extra constraint may be placed by
requiring that the ``flight time'' from M31 to Tucana's present
location is shorter than the Hubble time. We may estimate this by
assuming that Tucana is
a test particle
presently at the apocentre of a nearly
radial orbit, and integrating backwards in time to find when it was
propelled into such orbit. The estimate requires an assumption for the
gravitational potential of M31, for which we adopt a standard NFW halo
\citep{Navarro1997} with virial mass
$M_{200}=3\times 10^{12}\, M_\odot$ 
(\citealt{vanderMarel2012,Fardal2013},  about 3 times more
massive than the MW according to most current estimates, \citealt{Deason2019})
 and
concentration $c=7.8$, following \citet{Ludlow2016}\footnote{We choose 
such a setup for simplicity, as it is enough for our purposes here,
but acknowledge that the actual orbits of Tucana and M33 would look differently 
in detail when including a proper treatment for dynamical friction, the evolution of M31's
potential and the influence of an evolving cosmic web at early times.}.

\begin{figure}
\centering
\includegraphics[width=\linewidth]{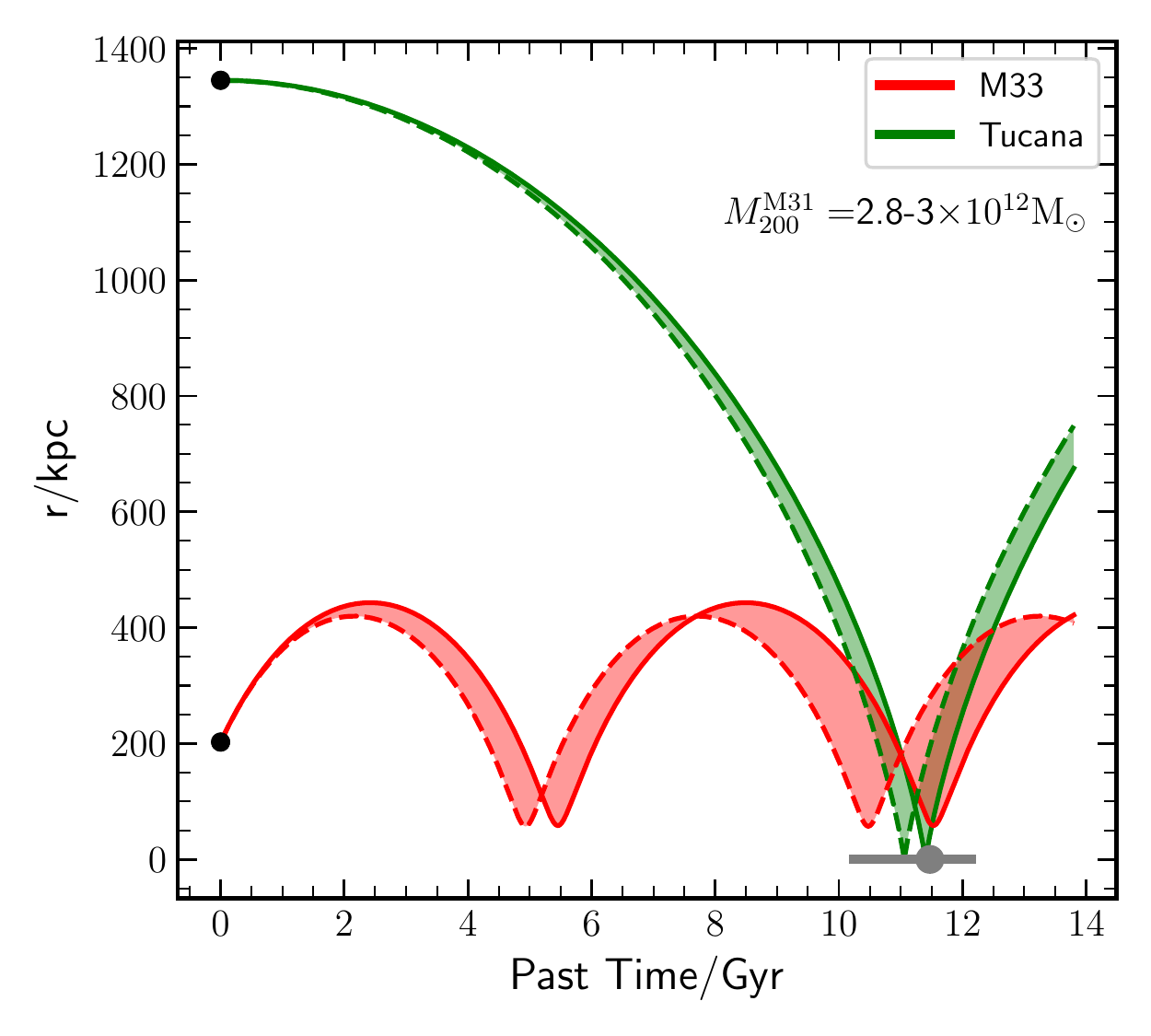}
\caption{Orbits of the Tucana and M33 dwarfs around M31, approximating M31's potential by an NFW potential with 
$M_{200}=2.8 (3) \times 10^{12}$ M$_\odot$ (shown in solid (dashed)
linestyles).   We assume Tucana is at present at the apocentre of a  radial orbit around M31. For M33 we employ its actual 3D velocity as derived from current observational data (see Tab.~\ref{tab:m31dataderived}).
Black points mark the dwarfs' distances at $z=0$.
The gray data point with errorbar indicates the obervational estimates for star formation cessation in Tucana dSph according to \citet{Monelli2010b}.  Specifically, the errorbar spans the temporal period between the build-up of $50-90\%$ of the stellar mass, and the point indicates when $70\%$ of the mass was acquired.
}
\label{fig:orbits}
\end{figure}

The orbit of Tucana, under these assumptions, is shown by the solid green curve in
Fig.~\ref{fig:orbits}. The dashed green curve assumes a different
$M_{200}$ of $2.8\times 10^{12}\, M_\odot$, and is
included just to illustrate the sensitivity of this result to variations
in M31's assumed virial mass. For these choices, we see that Tucana
could reach its present location if it was ejected from M31 roughly
$11$ Gyr ago. Remarkably, this roughly coincides with the time when Tucana ceased
forming stars, according to detailed modeling of its star formation
history by  \citet[][]{Monelli2010b}.

Finally, we may also integrate M33's orbit backwards assuming 
it is a test particle within
the same
M31 potential. The results are shown by the red curves in
Fig.~\ref{fig:orbits}, and suggest a further coincidence. M33 is today
approaching M31 on an orbit with a radial period of roughly $\sim 5$
Gyr. This places M33 near orbital pericentre at about the time when
Tucana may have been propelled into its highly-energetic orbit.

We note that this timing coincidence depends sensitively on the assumed
M31's mass, and would be much less compelling if M31's was, for
example, two times more (or less) massive than assumed here. Indeed,
for a virial mass as low as $\sim 10^{12}\, M_\odot$, M33's orbital
period would be so long that it could be on its first infall into M31
\citep[e.g.][]{Patel2017}. A virial mass that low, however, would make
M31's halo comparable to that of the Milky Way, which we find rather
unlikely given its much larger stellar mass. Note as well that our M33
integration assumes a static, spherical potential with no dynamical
friction, so the timing coincidence highlighted above may very well
disappear upon more detailed scrutiny.
Our main result, however (i.e., that there is enough time within a 
Hubble time for Tucana to travel to its current radial distance from M31),
remains valid.

In summary,
we believe that the sum of all these potential
coincidences (radial velocity, planar alignment, flight time and star
formation cessation, and concurrent pericentric passage) add up to a
credible case for a true physical association between Tucana and
M31/M33.

Although it may be difficult to prove such association beyond
reasonable doubt, it is important to identify what data may, in the
future, be used to validate or falsify this scenario.  Tighter
constraints on M31's virial mass, together with improved estimates of
M31 and M33's proper motions could help, as would estimates of
Tucana's proper motion. Indeed, the backsplash scenario posits that
Tucana is essentially at rest relative to M31, which allows us to
predict its proper motion: $\mu_{\rm RA*}=0.0206$ mas/yr and
$\mu_{\rm Dec}=-0.0754$ mas/yr, respectively.

Due to Tucana's large distance and lack of bright supergiant stars,
this measurement is probably beyond the reach of the Gaia satellite,
but it might be possible with HST and/or JWST \citep[see][and references therein]{McConnachie2021}.
 Confirming that
Tucana is indeed at rest relative to M31 would provide strong support
for the backsplash origin envisioned here.


\subsection{Tucana dSph as a satellite of M33}
We briefly discuss here the idea of Tucana as an ejected satellite 
of M33 in the context of M33’s predicted satellite population. 
Given M33’s  high stellar mass and implied halo mass, hierarchical clustering 
in LCDM predicts that it should have its own luminous satellites.
The satellite mass function (SMF) of dwarf galaxies is still observationally 
unconstrained. Nonetheless, one can utilize that of the MW --which is well 
known down to $M_*\approx 10^5$ M$_\odot$--, together with the assumption that the 
SMF is scale-free  \citep[similarly to  the underlying LCDM subhalo mass 
function,][]{Sales2013},  to quantify  M33’s expected satellite 
population. Considering the ratio of stellar masses between Tucana and M33\footnote{We adopt $M_*^{\rm Tuc}=5.6\times 10^6$M$_\odot$ and $M_*^{\rm M33}=2\times 10^9$M$_\odot$, computed by applying a mass-to-light ratio to the V-band luminosities in \citet{McConnachie2012}'s database. We assumed $M_*/L_V = 1$ for Tuc and $0.7$ for M33 \citep[see][]{Woo2008}.}, 
this exercise yields that M33 may have  harboured up to  $\sim5$ satellites with masses larger than 
that of Tucana.

As most likely the SMF is not scale-free and self-similar 
\citep[see e.g. ][]{Santos-Santos2022}, the actual number could in principle be lower. 
Indeed, the only example of a galaxy of similar mass as M33 with observed 
satellite candidates is the LMC, for which simulations and current observational data
constraints indicate that it may host up to $3$ satellites with $M_*>M_*^{\rm Tuc}$ 
\citep[see][]{Santos-Santos2021}.  Given that the LMC is now at first infall, 
this number probably represents the total number of original satellites of that 
mass expected around the LMC,  since there has not been enough 
time for MW tidal effects to disperse their orbits.
If, following the scenario we propose here,  M33 was accreted by M31 long 
ago,  even fewer satellites of M33 should remain at $z=0$ due to tidal stripping after
 subsequent pericentric passages \citep[see][]{Patel2018}.

 Therefore, a natural consequence of our proposed scenario for Tucana
 as a backsplash of an early-infalling M33 onto M31, is that M33 is
 likely to have lost its satellite population by now. This
 prediction agrees with the current observational data where only one
 satellite candidate is found within $\sim 100$ kpc around M33
 \citep[AndXXII,][]{Martin2009}.  

\section{Summary}\label{sec:conclu}

We have used the APOSTLE cosmological simulations to characterize the
population of galaxies dynamically associated with the two primary
galaxies (MW and M31) of the Local Group. ``Associated'' systems are
defined as those which have been, at some time during their evolution,
within the virial radius of one of the primaries. Associated
galaxies outside  the virial radius at $z=0$ are denoted as
``backsplash''; those inside $r_{200}$ are defined as
``satellites''. 

The fraction of dwarfs associated to a primary in APOSTLE drops
quickly outside its virial radius; from $\sim 50$\% at $400$ kpc
(roughly $2\times r_{200}$), to roughly $10\%$ at $600$ kpc.  The
most distant backsplash galaxy in all 4 APOSTLE volumes
analysed is located at $\sim 1.2$ Mpc, roughly 
$6\times$ the average virial radius of APOSTLE primaries.

Distant backsplash galaxies originate during the tidal disruption of
an accreted ``parent'' group of dwarfs, when they are propelled into
highly energetic orbits. Today
they are found mainly close to apocentre of nearly radial orbits (i.e., essentially
at rest) relative to their primaries, with a radial velocity
dispersion of only $\pm 29$ km/s beyond $\sim 600$ kpc.

We use this feature to examine which, if any, of the isolated Local
Group dwarfs could be a distant backsplash of the MW or M31. We focus,
in particular, on M31 backsplash candidates linked to the accretion of
M33, given the lack of obvious ``parent'' in the MW. (The Magellanic
Clouds are at present on first approach, and therefore could not have
caused backsplash systems as distant as the ones we examine here.)

There are at present eight LG dwarfs known outside $500$ kpc from the MW and
M31, and within $\sim 1.5$ Mpc from the Local Group midpoint:
Aquarius, Cetus, IC1613, LeoA, Sagittarius dIrr, Tucana, UGC4879 and
WLM. Several of these have low relative radial velocities relative to
M31, but one of them stands out: the Tucana dSph.

Tucana appears to be not only at rest relative to M31 in terms of its
radial velocity, but it also
lies almost perfectly on the orbital plane of M33 around M31. Further,
its flight time to its present location is roughly $\sim 10$ Gyr,
which coincides with the time when Tucana ceased forming stars. It
also coincides with one of the previous M33 pericentric passages
around M31, assuming that M31's virial mass is $\sim 3\times 10^{12}\,
M_\odot$. Each of these ``coincidences'' could be dismissed
individually, but, taken together, we believe that they make a
compelling case for identifying Tucana with a former satellite of
either M31 or M33 which was ejected from the M31 system, likely during
M33's first infall.

Further support for this scenario could come from tighter constraints
on the kinematics of M31 and M33 or on the virial mass of M31, or from
a measurement of Tucana's proper motion. For this scenario to work,
Tucana must be nearly at rest relative to M31, which allows us to
predict its proper motion:  $\mu_{\rm RA*}=0.0206$ mas/yr and
$\mu_{\rm Dec}=-0.0754$ mas/yr, respectively. Confirming such
prediction would provide strong evidence for the backsplash origin of
the Tucana dSph we propose here.



\section*{Acknowledgements}
We thank the referee for useful comments.
ISS acknowledges support
from the Arthur B. McDonald Canadian Astroparticle Physics
Research Institute and 
from the European Research Council (ERC) through Advanced Investigator
grant to C.S. Frenk, DMIDAS (GA 786910). 
We wish to acknowledge the generous contributions of all those
who made possible the Virgo Consortium’s EAGLE/APOSTLE
simulation projects. 
 This work used
the DiRAC@Durham facility managed by the Institute for Computational
Cosmology on behalf of the STFC DiRAC HPC Facility
(www.dirac.ac.uk). The equipment was funded by BEIS capital
funding via STFC capital grants ST/K00042X/1, ST/P002293/1,
ST/R002371/1 and ST/S002502/1, Durham University and STFC
operations grant ST/R000832/1. DiRAC is part of the National e-
Infrastructure.
This research made use of Astropy (http://www.astropy.org) a community-developed core Python package for Astronomy.

\section*{Data Availability}
The simulation data underlying this article can be shared on reasonable
request to the corresponding author.
The references for the observational data for Local Group dwarfs used in this article are listed in Sec.~\ref{sec:data}.

\bibliographystyle{mnras}
\bibliography{archive} 

\begin{thebibliography}{}
\makeatletter
\relax
\def\mn@urlcharsother{\let\do\@makeother \do\$\do\&\do\#\do\^\do\_\do\%\do\~}
\def\mn@doi{\begingroup\mn@urlcharsother \@ifnextchar [ {\mn@doi@}
  {\mn@doi@[]}}
\def\mn@doi@[#1]#2{\def\@tempa{#1}\ifx\@tempa\@empty \href
  {http://dx.doi.org/#2} {doi:#2}\else \href {http://dx.doi.org/#2} {#1}\fi
  \endgroup}
\def\mn@eprint#1#2{\mn@eprint@#1:#2::\@nil}
\def\mn@eprint@arXiv#1{\href {http://arxiv.org/abs/#1} {{\tt arXiv:#1}}}
\def\mn@eprint@dblp#1{\href {http://dblp.uni-trier.de/rec/bibtex/#1.xml}
  {dblp:#1}}
\def\mn@eprint@#1:#2:#3:#4\@nil{\def\@tempa {#1}\def\@tempb {#2}\def\@tempc
  {#3}\ifx \@tempc \@empty \let \@tempc \@tempb \let \@tempb \@tempa \fi \ifx
  \@tempb \@empty \def\@tempb {arXiv}\fi \@ifundefined
  {mn@eprint@\@tempb}{\@tempb:\@tempc}{\expandafter \expandafter \csname
  mn@eprint@\@tempb\endcsname \expandafter{\@tempc}}}

\bibitem[\protect\citeauthoryear{{Applebaum}, {Brooks}, {Christensen},
  {Munshi}, {Quinn}, {Shen}  \& {Tremmel}}{{Applebaum}
  et~al.}{2021}]{Applebaum2021}
{Applebaum} E.,  {Brooks} A.~M.,  {Christensen} C.~R.,  {Munshi} F.,  {Quinn}
  T.~R.,  {Shen} S.,   {Tremmel} M.,  2021, \mn@doi [\apj]
  {10.3847/1538-4357/abcafa}, \href
  {https://ui.adsabs.harvard.edu/abs/2021ApJ...906...96A} {906, 96}

\bibitem[\protect\citeauthoryear{{Bakels}, {Ludlow}  \& {Power}}{{Bakels}
  et~al.}{2021}]{Bakels2021}
{Bakels} L.,  {Ludlow} A.~D.,   {Power} C.,  2021, \mn@doi [\mnras]
  {10.1093/mnras/staa3979}, \href
  {https://ui.adsabs.harvard.edu/abs/2021MNRAS.501.5948B} {501, 5948}

\bibitem[\protect\citeauthoryear{{Balogh}, {Navarro}  \& {Morris}}{{Balogh}
  et~al.}{2000}]{Balogh2000}
{Balogh} M.~L.,  {Navarro} J.~F.,   {Morris} S.~L.,  2000, \mn@doi [\apj]
  {10.1086/309323}, \href
  {https://ui.adsabs.harvard.edu/abs/2000ApJ...540..113B} {540, 113}

\bibitem[\protect\citeauthoryear{{Barber}, {Starkenburg}, {Navarro},
  {McConnachie}  \& {Fattahi}}{{Barber} et~al.}{2014}]{Barber2014}
{Barber} C.,  {Starkenburg} E.,  {Navarro} J.~F.,  {McConnachie} A.~W.,
  {Fattahi} A.,  2014, \mn@doi [\mnras] {10.1093/mnras/stt1959}, \href
  {https://ui.adsabs.harvard.edu/abs/2014MNRAS.437..959B} {437, 959}

\bibitem[\protect\citeauthoryear{{Ben{\'\i}tez-Llambay}, {Navarro}, {Abadi},
  {Gottl{\"o}ber}, {Yepes}, {Hoffman}  \& {Steinmetz}}{{Ben{\'\i}tez-Llambay}
  et~al.}{2013}]{BenitezLlambay2013}
{Ben{\'\i}tez-Llambay} A.,  {Navarro} J.~F.,  {Abadi} M.~G.,  {Gottl{\"o}ber}
  S.,  {Yepes} G.,  {Hoffman} Y.,   {Steinmetz} M.,  2013, \mn@doi [\apjl]
  {10.1088/2041-8205/763/2/L41}, \href
  {https://ui.adsabs.harvard.edu/abs/2013ApJ...763L..41B} {763, L41}

\bibitem[\protect\citeauthoryear{{Besla}, {Kallivayalil}, {Hernquist}, {van der
  Marel}, {Cox}  \& {Kere{\v{s}}}}{{Besla} et~al.}{2012}]{Besla2012}
{Besla} G.,  {Kallivayalil} N.,  {Hernquist} L.,  {van der Marel} R.~P.,  {Cox}
  T.~J.,   {Kere{\v{s}}} D.,  2012, \mn@doi [\mnras]
  {10.1111/j.1365-2966.2012.20466.x}, \href
  {https://ui.adsabs.harvard.edu/abs/2012MNRAS.421.2109B} {421, 2109}

\bibitem[\protect\citeauthoryear{{Buck}, {Macci{\`o}}, {Dutton}, {Obreja}  \&
  {Frings}}{{Buck} et~al.}{2019}]{Buck2019}
{Buck} T.,  {Macci{\`o}} A.~V.,  {Dutton} A.~A.,  {Obreja} A.,   {Frings} J.,
  2019, \mn@doi [\mnras] {10.1093/mnras/sty2913}, \href
  {https://ui.adsabs.harvard.edu/abs/2019MNRAS.483.1314B} {483, 1314}

\bibitem[\protect\citeauthoryear{{Carignan}, {Chemin}, {Huchtmeier}  \&
  {Lockman}}{{Carignan} et~al.}{2006}]{Carignan2006}
{Carignan} C.,  {Chemin} L.,  {Huchtmeier} W.~K.,   {Lockman} F.~J.,  2006,
  \mn@doi [\apjl] {10.1086/503869}, \href
  {https://ui.adsabs.harvard.edu/abs/2006ApJ...641L.109C} {641, L109}

\bibitem[\protect\citeauthoryear{{Castellani}, {Marconi}  \&
  {Buonanno}}{{Castellani} et~al.}{1996}]{Castellani1996}
{Castellani} M.,  {Marconi} G.,   {Buonanno} R.,  1996, \aap, \href
  {https://ui.adsabs.harvard.edu/abs/1996A&A...310..715C} {310, 715}

\bibitem[\protect\citeauthoryear{{Crain} et~al.,}{{Crain}
  et~al.}{2015}]{Crain2015}
{Crain} R.~A.,  et~al., 2015, \mn@doi [\mnras] {10.1093/mnras/stv725}, \href
  {https://ui.adsabs.harvard.edu/abs/2015MNRAS.450.1937C} {450, 1937}

\bibitem[\protect\citeauthoryear{{Davis}, {Efstathiou}, {Frenk}  \&
  {White}}{{Davis} et~al.}{1985}]{Davis1985}
{Davis} M.,  {Efstathiou} G.,  {Frenk} C.~S.,   {White} S.~D.~M.,  1985,
  \mn@doi [\apj] {10.1086/163168}, \href
  {http://adsabs.harvard.edu/cgi-bin/nph-bib_query?bibcode=1985ApJ...292..371D&db_key=AST}
  {292, 371}

\bibitem[\protect\citeauthoryear{{Deason}, {Belokurov}  \& {Sanders}}{{Deason}
  et~al.}{2019}]{Deason2019}
{Deason} A.~J.,  {Belokurov} V.,   {Sanders} J.~L.,  2019, \mn@doi [\mnras]
  {10.1093/mnras/stz2793}, \href
  {https://ui.adsabs.harvard.edu/abs/2019MNRAS.490.3426D} {490, 3426}

\bibitem[\protect\citeauthoryear{{Dolag}, {Borgani}, {Murante}  \&
  {Springel}}{{Dolag} et~al.}{2009}]{Dolag2009}
{Dolag} K.,  {Borgani} S.,  {Murante} G.,   {Springel} V.,  2009, \mn@doi
  [\mnras] {10.1111/j.1365-2966.2009.15034.x}, \href
  {https://ui.adsabs.harvard.edu/abs/2009MNRAS.399..497D} {399, 497}

\bibitem[\protect\citeauthoryear{{Fardal} et~al.,}{{Fardal}
  et~al.}{2013}]{Fardal2013}
{Fardal} M.~A.,  et~al., 2013, \mn@doi [\mnras] {10.1093/mnras/stt1121}, \href
  {https://ui.adsabs.harvard.edu/abs/2013MNRAS.434.2779F} {434, 2779}

\bibitem[\protect\citeauthoryear{{Fattahi} et~al.,}{{Fattahi}
  et~al.}{2016}]{Fattahi2016}
{Fattahi} A.,  et~al., 2016, \mn@doi [\mnras] {10.1093/mnras/stv2970}, \href
  {https://ui.adsabs.harvard.edu/abs/2016MNRAS.457..844F} {457, 844}

\bibitem[\protect\citeauthoryear{{Fattahi}, {Navarro}, {Frenk}, {Oman},
  {Sawala}  \& {Schaller}}{{Fattahi} et~al.}{2018}]{Fattahi2018}
{Fattahi} A.,  {Navarro} J.~F.,  {Frenk} C.~S.,  {Oman} K.~A.,  {Sawala} T.,
  {Schaller} M.,  2018, \mn@doi [\mnras] {10.1093/mnras/sty408}, \href
  {https://ui.adsabs.harvard.edu/abs/2018MNRAS.476.3816F} {476, 3816}

\bibitem[\protect\citeauthoryear{{Fraternali}, {Tolstoy}, {Irwin}  \&
  {Cole}}{{Fraternali} et~al.}{2009}]{Fraternali2009}
{Fraternali} F.,  {Tolstoy} E.,  {Irwin} M.~J.,   {Cole} A.~A.,  2009, \mn@doi
  [\aap] {10.1051/0004-6361/200810830}, \href
  {https://ui.adsabs.harvard.edu/abs/2009A&A...499..121F} {499, 121}

\bibitem[\protect\citeauthoryear{{Garrison-Kimmel}, {Boylan-Kolchin}, {Bullock}
   \& {Lee}}{{Garrison-Kimmel} et~al.}{2014}]{Garrison-Kimmel2014}
{Garrison-Kimmel} S.,  {Boylan-Kolchin} M.,  {Bullock} J.~S.,   {Lee} K.,
  2014, \mn@doi [\mnras] {10.1093/mnras/stt2377}, \href
  {https://ui.adsabs.harvard.edu/abs/2014MNRAS.438.2578G} {438, 2578}

\bibitem[\protect\citeauthoryear{{Geha}, {Blanton}, {Yan}  \& {Tinker}}{{Geha}
  et~al.}{2012}]{Geha2012}
{Geha} M.,  {Blanton} M.~R.,  {Yan} R.,   {Tinker} J.~L.,  2012, \mn@doi [\apj]
  {10.1088/0004-637X/757/1/85}, \href
  {https://ui.adsabs.harvard.edu/abs/2012ApJ...757...85G} {757, 85}

\bibitem[\protect\citeauthoryear{{Gill}, {Knebe}  \& {Gibson}}{{Gill}
  et~al.}{2005}]{Gill2005}
{Gill} S. P.~D.,  {Knebe} A.,   {Gibson} B.~K.,  2005, \mn@doi [\mnras]
  {10.1111/j.1365-2966.2004.08562.x}, \href
  {https://ui.adsabs.harvard.edu/abs/2005MNRAS.356.1327G} {356, 1327}

\bibitem[\protect\citeauthoryear{{Grebel}}{{Grebel}}{1998}]{Grebel1998}
{Grebel} E.~K.,  1998, Highlights of Astronomy, \href
  {https://ui.adsabs.harvard.edu/abs/1998HiA....11..125G} {11A, 125}

\bibitem[\protect\citeauthoryear{{Kallivayalil}, {van der Marel}, {Besla},
  {Anderson}  \& {Alcock}}{{Kallivayalil} et~al.}{2013}]{Kallivayalil2013}
{Kallivayalil} N.,  {van der Marel} R.~P.,  {Besla} G.,  {Anderson} J.,
  {Alcock} C.,  2013, \mn@doi [\apj] {10.1088/0004-637X/764/2/161}, \href
  {https://ui.adsabs.harvard.edu/abs/2013ApJ...764..161K} {764, 161}

\bibitem[\protect\citeauthoryear{{Karachentsev} et~al.,}{{Karachentsev}
  et~al.}{2001}]{Karachentsev2001}
{Karachentsev} I.~D.,  et~al., 2001, \mn@doi [\aap]
  {10.1051/0004-6361:20011344}, \href
  {https://ui.adsabs.harvard.edu/abs/2001A&A...379..407K} {379, 407}

\bibitem[\protect\citeauthoryear{{Karachentsev}, {Kniazev}  \&
  {Sharina}}{{Karachentsev} et~al.}{2015}]{Karachentsev2015}
{Karachentsev} I.~D.,  {Kniazev} A.~Y.,   {Sharina} M.~E.,  2015, \mn@doi
  [Astronomische Nachrichten] {10.1002/asna.201512207}, \href
  {https://ui.adsabs.harvard.edu/abs/2015AN....336..707K} {336, 707}

\bibitem[\protect\citeauthoryear{{Knebe}, {Libeskind}, {Knollmann},
  {Martinez-Vaquero}, {Yepes}, {Gottl{\"o}ber}  \& {Hoffman}}{{Knebe}
  et~al.}{2011}]{Knebe2011a}
{Knebe} A.,  {Libeskind} N.~I.,  {Knollmann} S.~R.,  {Martinez-Vaquero} L.~A.,
  {Yepes} G.,  {Gottl{\"o}ber} S.,   {Hoffman} Y.,  2011, \mn@doi [\mnras]
  {10.1111/j.1365-2966.2010.17924.x}, \href
  {https://ui.adsabs.harvard.edu/abs/2011MNRAS.412..529K} {412, 529}

\bibitem[\protect\citeauthoryear{{Komatsu} et~al.,}{{Komatsu}
  et~al.}{2011}]{Komatsu2011}
{Komatsu} E.,  et~al., 2011, \mn@doi [\apjs] {10.1088/0067-0049/192/2/18},
  \href {http://adsabs.harvard.edu/abs/2011ApJS..192...18K} {192, 18}

\bibitem[\protect\citeauthoryear{{Laporte}, {Johnston}, {G{\'o}mez},
  {Garavito-Camargo}  \& {Besla}}{{Laporte} et~al.}{2018}]{Laporte2018b}
{Laporte} C. F.~P.,  {Johnston} K.~V.,  {G{\'o}mez} F.~A.,  {Garavito-Camargo}
  N.,   {Besla} G.,  2018, \mn@doi [\mnras] {10.1093/mnras/sty1574}, \href
  {https://ui.adsabs.harvard.edu/abs/2018MNRAS.481..286L} {481, 286}

\bibitem[\protect\citeauthoryear{{Lavery} \& {Mighell}}{{Lavery} \&
  {Mighell}}{1992}]{Lavery1992}
{Lavery} R.~J.,  {Mighell} K.~J.,  1992, \mn@doi [\aj] {10.1086/116042}, \href
  {https://ui.adsabs.harvard.edu/abs/1992AJ....103...81L} {103, 81}

\bibitem[\protect\citeauthoryear{{Ludlow}, {Navarro}, {Springel}, {Jenkins},
  {Frenk}  \& {Helmi}}{{Ludlow} et~al.}{2009}]{Ludlow2009}
{Ludlow} A.~D.,  {Navarro} J.~F.,  {Springel} V.,  {Jenkins} A.,  {Frenk}
  C.~S.,   {Helmi} A.,  2009, \mn@doi [\apj] {10.1088/0004-637X/692/1/931},
  \href {https://ui.adsabs.harvard.edu/abs/2009ApJ...692..931L} {692, 931}

\bibitem[\protect\citeauthoryear{{Ludlow}, {Bose}, {Angulo}, {Wang},
  {Hellwing}, {Navarro}, {Cole}  \& {Frenk}}{{Ludlow}
  et~al.}{2016}]{Ludlow2016}
{Ludlow} A.~D.,  {Bose} S.,  {Angulo} R.~E.,  {Wang} L.,  {Hellwing} W.~A.,
  {Navarro} J.~F.,  {Cole} S.,   {Frenk} C.~S.,  2016, \mn@doi [\mnras]
  {10.1093/mnras/stw1046}, \href
  {https://ui.adsabs.harvard.edu/abs/2016MNRAS.460.1214L} {460, 1214}

\bibitem[\protect\citeauthoryear{{Makarova}, {Makarov}, {Karachentsev}, {Tully}
   \& {Rizzi}}{{Makarova} et~al.}{2017}]{Makarova2017}
{Makarova} L.~N.,  {Makarov} D.~I.,  {Karachentsev} I.~D.,  {Tully} R.~B.,
  {Rizzi} L.,  2017, \mn@doi [\mnras] {10.1093/mnras/stw2502}, \href
  {https://ui.adsabs.harvard.edu/abs/2017MNRAS.464.2281M} {464, 2281}

\bibitem[\protect\citeauthoryear{{Mamon}, {Sanchis}, {Salvador-Sol{\'e}}  \&
  {Solanes}}{{Mamon} et~al.}{2004}]{Mamon2004}
{Mamon} G.~A.,  {Sanchis} T.,  {Salvador-Sol{\'e}} E.,   {Solanes} J.~M.,
  2004, \mn@doi [\aap] {10.1051/0004-6361:20034155}, \href
  {https://ui.adsabs.harvard.edu/abs/2004A&A...414..445M} {414, 445}

\bibitem[\protect\citeauthoryear{{Martin} et~al.,}{{Martin}
  et~al.}{2009}]{Martin2009}
{Martin} N.~F.,  et~al., 2009, \mn@doi [\apj] {10.1088/0004-637X/705/1/758},
  \href {https://ui.adsabs.harvard.edu/abs/2009ApJ...705..758M} {705, 758}

\bibitem[\protect\citeauthoryear{{McConnachie}}{{McConnachie}}{2012}]{McConnachie2012}
{McConnachie} A.~W.,  2012, \mn@doi [\aj] {10.1088/0004-6256/144/1/4}, \href
  {https://ui.adsabs.harvard.edu/abs/2012AJ....144....4M} {144, 4}

\bibitem[\protect\citeauthoryear{{McConnachie} et~al.,}{{McConnachie}
  et~al.}{2008}]{McConnachie2008}
{McConnachie} A.~W.,  et~al., 2008, \mn@doi [\apj] {10.1086/591313}, \href
  {https://ui.adsabs.harvard.edu/abs/2008ApJ...688.1009M} {688, 1009}

\bibitem[\protect\citeauthoryear{{McConnachie} et~al.,}{{McConnachie}
  et~al.}{2009}]{McConnachie2009}
{McConnachie} A.~W.,  et~al., 2009, \mn@doi [\nat] {10.1038/nature08327}, \href
  {https://ui.adsabs.harvard.edu/abs/2009Natur.461...66M} {461, 66}

\bibitem[\protect\citeauthoryear{{McConnachie}, {Higgs}, {Thomas}, {Venn},
  {C{\^o}t{\'e}}, {Battaglia}  \& {Lewis}}{{McConnachie}
  et~al.}{2021}]{McConnachie2021}
{McConnachie} A.~W.,  {Higgs} C.~R.,  {Thomas} G.~F.,  {Venn} K.~A.,
  {C{\^o}t{\'e}} P.,  {Battaglia} G.,   {Lewis} G.~F.,  2021, \mn@doi [\mnras]
  {10.1093/mnras/staa3740}, \href
  {https://ui.adsabs.harvard.edu/abs/2021MNRAS.501.2363M} {501, 2363}

\bibitem[\protect\citeauthoryear{{McMillan}}{{McMillan}}{2011}]{McMillan2011}
{McMillan} P.~J.,  2011, \mn@doi [\mnras] {10.1111/j.1365-2966.2011.18564.x},
  \href {https://ui.adsabs.harvard.edu/abs/2011MNRAS.414.2446M} {414, 2446}

\bibitem[\protect\citeauthoryear{{Monelli} et~al.,}{{Monelli}
  et~al.}{2010a}]{Monelli2010a}
{Monelli} M.,  et~al., 2010a, \mn@doi [\apj] {10.1088/0004-637X/720/2/1225},
  \href {https://ui.adsabs.harvard.edu/abs/2010ApJ...720.1225M} {720, 1225}

\bibitem[\protect\citeauthoryear{{Monelli} et~al.,}{{Monelli}
  et~al.}{2010b}]{Monelli2010b}
{Monelli} M.,  et~al., 2010b, \mn@doi [\apj] {10.1088/0004-637X/722/2/1864},
  \href {https://ui.adsabs.harvard.edu/abs/2010ApJ...722.1864M} {722, 1864}

\bibitem[\protect\citeauthoryear{{Navarro}, {Frenk}  \& {White}}{{Navarro}
  et~al.}{1997}]{Navarro1997}
{Navarro} J.~F.,  {Frenk} C.~S.,   {White} S.~D.~M.,  1997, \apj, \href
  {http://adsabs.harvard.edu/abs/1997ApJ...490..493N} {490, 493}

\bibitem[\protect\citeauthoryear{{Newton} et~al.,}{{Newton}
  et~al.}{2021}]{Newton2021}
{Newton} O.,  et~al., 2021, arXiv e-prints, \href
  {https://ui.adsabs.harvard.edu/abs/2021arXiv210411242N} {p. arXiv:2104.11242}

\bibitem[\protect\citeauthoryear{{Patel}, {Besla}  \& {Sohn}}{{Patel}
  et~al.}{2017}]{Patel2017}
{Patel} E.,  {Besla} G.,   {Sohn} S.~T.,  2017, \mn@doi [\mnras]
  {10.1093/mnras/stw2616}, \href
  {https://ui.adsabs.harvard.edu/abs/2017MNRAS.464.3825P} {464, 3825}

\bibitem[\protect\citeauthoryear{{Patel}, {Carlin}, {Tollerud}, {Collins}  \&
  {Dooley}}{{Patel} et~al.}{2018}]{Patel2018}
{Patel} E.,  {Carlin} J.~L.,  {Tollerud} E.~J.,  {Collins} M. L.~M.,   {Dooley}
  G.~A.,  2018, \mn@doi [\mnras] {10.1093/mnras/sty1946}, \href
  {https://ui.adsabs.harvard.edu/abs/2018MNRAS.480.1883P} {480, 1883}

\bibitem[\protect\citeauthoryear{{Pereira Wilson}, {Navarro}, {Santos Santos}
  \& {Benitez Llambay}}{{Pereira Wilson} et~al.}{2022}]{PereiraWilson2022}
{Pereira Wilson} M.,  {Navarro} J.,  {Santos Santos} I.,   {Benitez Llambay}
  A.,  2022, arXiv e-prints, \href
  {https://ui.adsabs.harvard.edu/abs/2022arXiv220605338P} {p. arXiv:2206.05338}

\bibitem[\protect\citeauthoryear{{Polzin}, {van Dokkum}, {Danieli}, {Greco}  \&
  {Romanowsky}}{{Polzin} et~al.}{2021}]{Polzin2021}
{Polzin} A.,  {van Dokkum} P.,  {Danieli} S.,  {Greco} J.~P.,   {Romanowsky}
  A.~J.,  2021, \mn@doi [\apjl] {10.3847/2041-8213/ac024f}, \href
  {https://ui.adsabs.harvard.edu/abs/2021ApJ...914L..23P} {914, L23}

\bibitem[\protect\citeauthoryear{{Sales}, {Navarro}, {Abadi}  \&
  {Steinmetz}}{{Sales} et~al.}{2007}]{Sales2007}
{Sales} L.~V.,  {Navarro} J.~F.,  {Abadi} M.~G.,   {Steinmetz} M.,  2007,
  \mn@doi [\mnras] {10.1111/j.1365-2966.2007.12026.x}, \href
  {https://ui.adsabs.harvard.edu/abs/2007MNRAS.379.1475S} {379, 1475}

\bibitem[\protect\citeauthoryear{{Sales}, {Navarro}, {Cooper}, {White}, {Frenk}
   \& {Helmi}}{{Sales} et~al.}{2011}]{Sales2011}
{Sales} L.~V.,  {Navarro} J.~F.,  {Cooper} A.~P.,  {White} S. D.~M.,  {Frenk}
  C.~S.,   {Helmi} A.,  2011, \mn@doi [\mnras]
  {10.1111/j.1365-2966.2011.19514.x}, \href
  {https://ui.adsabs.harvard.edu/abs/2011MNRAS.418..648S} {418, 648}

\bibitem[\protect\citeauthoryear{{Sales}, {Wang}, {White}  \&
  {Navarro}}{{Sales} et~al.}{2013}]{Sales2013}
{Sales} L.~V.,  {Wang} W.,  {White} S. D.~M.,   {Navarro} J.~F.,  2013, \mn@doi
  [\mnras] {10.1093/mnras/sts054}, \href
  {https://ui.adsabs.harvard.edu/abs/2013MNRAS.428..573S} {428, 573}

\bibitem[\protect\citeauthoryear{{Sand} et~al.,}{{Sand}
  et~al.}{2022}]{Sand2022}
{Sand} D.~J.,  et~al., 2022, arXiv e-prints, \href
  {https://ui.adsabs.harvard.edu/abs/2022arXiv220509129S} {p. arXiv:2205.09129}

\bibitem[\protect\citeauthoryear{{Santos-Santos}, {Fattahi}, {Sales}  \&
  {Navarro}}{{Santos-Santos} et~al.}{2021}]{Santos-Santos2021}
{Santos-Santos} I. M.~E.,  {Fattahi} A.,  {Sales} L.~V.,   {Navarro} J.~F.,
  2021, \mn@doi [\mnras] {10.1093/mnras/stab1020}, \href
  {https://ui.adsabs.harvard.edu/abs/2021MNRAS.504.4551S} {504, 4551}

\bibitem[\protect\citeauthoryear{{Santos-Santos}, {Sales}, {Fattahi}  \&
  {Navarro}}{{Santos-Santos} et~al.}{2022}]{Santos-Santos2022}
{Santos-Santos} I. M.~E.,  {Sales} L.~V.,  {Fattahi} A.,   {Navarro} J.~F.,
  2022, \mn@doi [\mnras] {10.1093/mnras/stac2057}, \href
  {https://ui.adsabs.harvard.edu/abs/2022MNRAS.515.3685S} {515, 3685}

\bibitem[\protect\citeauthoryear{{Savino}, {Tolstoy}, {Salaris}, {Monelli}  \&
  {de Boer}}{{Savino} et~al.}{2019}]{Savino2019}
{Savino} A.,  {Tolstoy} E.,  {Salaris} M.,  {Monelli} M.,   {de Boer} T.~J.~L.,
   2019, \mn@doi [\aap] {10.1051/0004-6361/201936077}, \href
  {https://ui.adsabs.harvard.edu/abs/2019A&A...630A.116S} {630, A116}

\bibitem[\protect\citeauthoryear{{Savino} et~al.,}{{Savino}
  et~al.}{2022}]{Savino2022}
{Savino} A.,  et~al., 2022, \mn@doi [\apj] {10.3847/1538-4357/ac91cb}, \href
  {https://ui.adsabs.harvard.edu/abs/2022ApJ...938..101S} {938, 101}

\bibitem[\protect\citeauthoryear{{Sawala} et~al.,}{{Sawala}
  et~al.}{2016}]{Sawala2016}
{Sawala} T.,  et~al., 2016, \mn@doi [\mnras] {10.1093/mnras/stw145}, \href
  {https://ui.adsabs.harvard.edu/abs/2016MNRAS.457.1931S} {457, 1931}

\bibitem[\protect\citeauthoryear{{Schaye} et~al.,}{{Schaye}
  et~al.}{2015}]{Schaye2015}
{Schaye} J.,  et~al., 2015, \mn@doi [\mnras] {10.1093/mnras/stu2058}, \href
  {http://adsabs.harvard.edu/abs/2015MNRAS.446..521S} {446, 521}

\bibitem[\protect\citeauthoryear{{Sch{\"o}nrich}, {Binney}  \&
  {Dehnen}}{{Sch{\"o}nrich} et~al.}{2010}]{Schonrich2010}
{Sch{\"o}nrich} R.,  {Binney} J.,   {Dehnen} W.,  2010, \mn@doi [\mnras]
  {10.1111/j.1365-2966.2010.16253.x}, \href
  {https://ui.adsabs.harvard.edu/abs/2010MNRAS.403.1829S} {403, 1829}

\bibitem[\protect\citeauthoryear{{Simpson}, {Grand}, {G{\'o}mez}, {Marinacci},
  {Pakmor}, {Springel}, {Campbell}  \& {Frenk}}{{Simpson}
  et~al.}{2018}]{Simpson2018}
{Simpson} C.~M.,  {Grand} R. J.~J.,  {G{\'o}mez} F.~A.,  {Marinacci} F.,
  {Pakmor} R.,  {Springel} V.,  {Campbell} D. J.~R.,   {Frenk} C.~S.,  2018,
  \mn@doi [\mnras] {10.1093/mnras/sty774}, \href
  {https://ui.adsabs.harvard.edu/abs/2018MNRAS.478..548S} {478, 548}

\bibitem[\protect\citeauthoryear{{Springel}, {Yoshida}  \& {White}}{{Springel}
  et~al.}{2001}]{Springel2001}
{Springel} V.,  {Yoshida} N.,   {White} S. D.~M.,  2001, \mn@doi [\na]
  {10.1016/S1384-1076(01)00042-2}, \href
  {https://ui.adsabs.harvard.edu/abs/2001NewA....6...79S} {6, 79}

\bibitem[\protect\citeauthoryear{{Taibi}, {Battaglia}, {Rejkuba}, {Leaman},
  {Kacharov}, {Iorio}, {Jablonka}  \& {Zoccali}}{{Taibi}
  et~al.}{2020}]{Taibi2020}
{Taibi} S.,  {Battaglia} G.,  {Rejkuba} M.,  {Leaman} R.,  {Kacharov} N.,
  {Iorio} G.,  {Jablonka} P.,   {Zoccali} M.,  2020, \mn@doi [\aap]
  {10.1051/0004-6361/201937240}, \href
  {https://ui.adsabs.harvard.edu/abs/2020A&A...635A.152T} {635, A152}

\bibitem[\protect\citeauthoryear{{Teyssier}, {Johnston}  \&
  {Kuhlen}}{{Teyssier} et~al.}{2012}]{Teyssier2012}
{Teyssier} M.,  {Johnston} K.~V.,   {Kuhlen} M.,  2012, \mn@doi [\mnras]
  {10.1111/j.1365-2966.2012.21793.x}, \href
  {https://ui.adsabs.harvard.edu/abs/2012MNRAS.426.1808T} {426, 1808}

\bibitem[\protect\citeauthoryear{{Tolstoy}, {Hill}  \& {Tosi}}{{Tolstoy}
  et~al.}{2009}]{Tolstoy2009}
{Tolstoy} E.,  {Hill} V.,   {Tosi} M.,  2009, \mn@doi [\araa]
  {10.1146/annurev-astro-082708-101650}, \href
  {https://ui.adsabs.harvard.edu/abs/2009ARA&A..47..371T} {47, 371}

\bibitem[\protect\citeauthoryear{{Wang} et~al.,}{{Wang}
  et~al.}{2011}]{Wang2011}
{Wang} J.,  et~al., 2011, \mn@doi [\mnras] {10.1111/j.1365-2966.2011.18220.x},
  \href {https://ui.adsabs.harvard.edu/abs/2011MNRAS.413.1373W} {413, 1373}

\bibitem[\protect\citeauthoryear{{Weisz}, {Dolphin}, {Skillman}, {Holtzman},
  {Gilbert}, {Dalcanton}  \& {Williams}}{{Weisz} et~al.}{2014}]{Weisz2014}
{Weisz} D.~R.,  {Dolphin} A.~E.,  {Skillman} E.~D.,  {Holtzman} J.,  {Gilbert}
  K.~M.,  {Dalcanton} J.~J.,   {Williams} B.~F.,  2014, \mn@doi [\apj]
  {10.1088/0004-637X/789/2/147}, \href
  {http://adsabs.harvard.edu/abs/2014ApJ...789..147W} {789, 147}

\bibitem[\protect\citeauthoryear{{Whiting}, {Hau}  \& {Irwin}}{{Whiting}
  et~al.}{1999}]{Whiting1999}
{Whiting} A.~B.,  {Hau} G. K.~T.,   {Irwin} M.,  1999, \mn@doi [\aj]
  {10.1086/301142}, \href
  {https://ui.adsabs.harvard.edu/abs/1999AJ....118.2767W} {118, 2767}

\bibitem[\protect\citeauthoryear{{Woo}, {Courteau}  \& {Dekel}}{{Woo}
  et~al.}{2008}]{Woo2008}
{Woo} J.,  {Courteau} S.,   {Dekel} A.,  2008, \mn@doi [\mnras]
  {10.1111/j.1365-2966.2008.13770.x}, \href
  {https://ui.adsabs.harvard.edu/abs/2008MNRAS.390.1453W} {390, 1453}

\bibitem[\protect\citeauthoryear{{van der Marel}, {Fardal}, {Besla}, {Beaton},
  {Sohn}, {Anderson}, {Brown}  \& {Guhathakurta}}{{van der Marel}
  et~al.}{2012}]{vanderMarel2012}
{van der Marel} R.~P.,  {Fardal} M.,  {Besla} G.,  {Beaton} R.~L.,  {Sohn}
  S.~T.,  {Anderson} J.,  {Brown} T.,   {Guhathakurta} P.,  2012, \mn@doi
  [\apj] {10.1088/0004-637X/753/1/8}, \href
  {https://ui.adsabs.harvard.edu/abs/2012ApJ...753....8V} {753, 8}

\bibitem[\protect\citeauthoryear{{van der Marel}, {Fardal}, {Sohn}, {Patel},
  {Besla}, {del Pino}, {Sahlmann}  \& {Watkins}}{{van der Marel}
  et~al.}{2019}]{vanderMarel2019}
{van der Marel} R.~P.,  {Fardal} M.~A.,  {Sohn} S.~T.,  {Patel} E.,  {Besla}
  G.,  {del Pino} A.,  {Sahlmann} J.,   {Watkins} L.~L.,  2019, \mn@doi [\apj]
  {10.3847/1538-4357/ab001b}, \href
  {https://ui.adsabs.harvard.edu/abs/2019ApJ...872...24V} {872, 24}

\makeatother
\end{thebibliography}



\bsp	
\label{lastpage}

\end{document}